\documentclass[pre,aps,twocolumn,superscriptaddress]{revtex4}
\pdfoutput=1
\usepackage{amssymb}
\usepackage{epsfig}
\usepackage{amsmath}
\usepackage{subfigure}
\usepackage{graphicx}
\usepackage{textcomp}
\usepackage{url}
\usepackage{float}
\usepackage{color}
\usepackage{cases}
\usepackage[normalem]{ulem}

\usepackage[titletoc,title]{appendix}

\usepackage[colorlinks=true, urlcolor=blue, anchorcolor=blue, citecolor=blue,filecolor=blue,linkcolor=blue,menucolor=blue
]{hyperref}

\begin{document}
\title{Geometrical optics of first-passage functionals of random acceleration}

\author{Baruch Meerson}
\email{meerson@mail.huji.ac.il}
\affiliation{Racah Institute of Physics, Hebrew University of
Jerusalem, Jerusalem 91904, Israel}

\begin{abstract}
Random acceleration is a fundamental stochastic process encountered in many applications. In the one-dimensional version of the process a particle is randomly accelerated according to the Langevin equation $\ddot{x}(t) = \sqrt{2D} \xi(t)$, where $x(t)$ is the particle's coordinate,
$\xi(t)$ is Gaussian white noise with zero mean, and $D$ is the particle velocity diffusion constant. Here we evaluate the $A\to 0$ tail of the distribution $P_n(A|L)$ of the functional $I[x(t)]=\int_0^{T} x^n(t) dt=A$,
where $T$ is the first-passage time of the particle from a specified point $x=L$ to the origin, and $n\geq 0$.   We employ the optimal fluctuation method akin to geometrical optics. Its crucial element is determination of the optimal path -- the most probable realization of the random acceleration process $x(t)$, conditioned on specified $A$, $n$ and $L$. The optimal path dominates the $A\to 0$ tail of $P_n(A|L)$. We show that this tail has a universal essential singularity, $P_n(A\to 0|L) \sim \exp\left(-\frac{\alpha_n L^{3n+2}}{DA^3}\right)$, where $\alpha_n$ is an $n$-dependent number which we calculate analytically for $n=0,1$ and $2$ and numerically for other $n$. For $n=0$ our result agrees with the asymptotic of the previously found first-passage time distribution.

\end{abstract}

\maketitle
\nopagebreak
\section{Introduction}
\label{intro}

The random acceleration process is governed by the Langevin equation
\begin{equation}\label{Langevin}
\ddot{x}(t) = \sqrt{2D} \xi(t)\,.
\end{equation}
This equation describes  the position of a particle moving along the $x$-axis and subject to a random force
which is modeled as a Gaussian white noise with zero mean, $\langle \xi(t) \xi(t')\rangle =\delta (t-t')$. Alternatively, $x(t)$ can be considered as the integral of a Brownian motion over time. The random acceleration is a fundamental stochastic process in its own right. On the one hand, it serves as a simple example of a non-Markovian process (which becomes Markovian when considered in two dimensions $x$ and $\dot{x}$, see \textit{e.g.} Ref. \cite{BMS2013}).  On the other hand, its mathematical equivalents have found a variety of applications in physics: from a simplified description of free semiflexible polymer chains in narrow channels \cite{Burkhardt1993,Burkhardt1997,Burkhardt2001,Burkhardt2007,Burkhardt2014,polymerexplain} to interface growth in 1+1 dimensions \cite{GB1991,Majumdar2001,BS} and to decaying turbulence in the Burgers equation \cite{Sinai,Velageas}. In all these systems it is a spatial coordinate which plays the role of time $t$ in Eq.~(\ref{Langevin}), while the polymer shape, or the interface shape, \textit{etc.} plays the role of $x$.

Here we are interested in the statistics of first-passage functionals  of the form $I[x(t)]=\int_0^T x^n(t)\,dt$, defined up to the time of first passage time $T$ of the process, starting say
at $x=L>0$, to a specified point in space, for example to the origin.  The case $n=0$ corresponds to the statistics of the first-passage time itself. The case $n=1$ corresponds to the area under the graph of $x(t)$ until the first passage to the origin. In the context of interface growth, governed by the noisy Mullins-Herring equation \cite{GB1991,BS}, it describes the area under the stochastic interface until it crosses a zero level in space for the first time. The case $n=2$ corresponds to the statistics of the moment of inertia of a semiflexible polymer chain of a given length in narrow channels. It is natural then to attempt to calculate the distribution of the values of the first-passage functional $I[x(t)]=\int_0^T x^n(t)\,dt$  for arbitrary $n$.

For comparison, the statistics of first-passage Brownian functionals \cite{Kearney,Majumdarreview} -- where $x(t)$ is a Brownian motion -- is well studied, see Ref. \cite{MM2020} and references therein. For the random acceleration process, however, the problem has been solved only for $n=0$, that is only for the statistics of the first-passage time itself \cite{McKean,Marshall,Burkhardt2014}. In the absence of general results for the complete distribution $P_n(A|L)$ of the values $I[x(t)]=\int_0^T x^n(t)\,dt=A$, here we focus on the $A\to 0$ tail of this distribution. We show that this tail exhibits an essential singularity, see Eq.~(\ref{Pscaling}) below. To achieve this goal,  we employ the optimal fluctuation method akin to geometrical optics \cite{geometricaloptics}. The  method relies on the determination of
the optimal path, that is the most likely realization of the process $x(t)$, conditioned on the specified value of $A\to 0$ at given $n$ and $L$. It is this optimal path that dominates the $A\to 0$ tail of $P_n(A|L)$. Previously, the geometrical optics was applied to a plethora of problems related to statistics of Brownian motion \cite{Grosberg,Ikeda,Schuss,SM2019,M2019,MS2019a,M2019b,MM2020,Agranovetal,M2020}.  An extension of the method to the random acceleration is a natural next step.

Here is a plan of the remainder of the paper. We complete the formulation of
the problem, establish the scaling properties of $P_n(A\to 0|L)$ and derive the governing equation of the optimal fluctuation method in Sec. \ref{govern}. Some analytical and numerical solutions for different $n$ are presented in Sec. \ref{solution}.
Section~\ref{discussion} includes a brief summary and an extension of our results. A technical derivation is delegated to the Appendix.

\section{Formulation of the problem and governing equations}
\label{govern}

We start by completing the formulation of the problem. The initial and final positions of the particle are
\begin{equation}\label{BC1}
x(t=0)=L\,,\quad x(T) = 0\,,
\end{equation}
where $T$ is the first passage time to the origin, and $L$ can be assumed positive without loss of generality.  We assume for simplicity that the particle starts with zero velocity:
\begin{equation}\label{BC2}
\dot{x}(t=0) = 0\,.
\end{equation}
We consider  first-passage functionals of the form $I[x(t)]=\int_0^T x^n(t)\,dt$ and study the probability distribution $P_n(A|L)$ of their values $A$:
\begin{equation}\label{constraint0}
\int_0^T x^n(t)\,dt = A\,.
\end{equation}
Equations~(\ref{Langevin})-(\ref{constraint0}) define the stochastic problem completely. Their dimensional analysis (notice that the units of $A$ depend on $n$) yields \cite{scalinganalysis} the following \emph{exact} scaling behavior of $P_n(A|L)$:
\begin{equation}\label{scaling}
P_n(A|L) = \frac{D^{1/3}}{L^{n+\frac{2}{3}}}\,F_n\left(\frac{D^{1/3} A}{L^{n+\frac{2}{3}}}\right)\,,
\end{equation}
with a dimensionless scaling function $F_n(z)$ of the dimensionless argument $z=D^{1/3} A L^{-n-\frac{2}{3}}$.  The scaling function $F_n(z)$ is presently unknown. On the physical grounds this function is expected to have a single maximum, of order $1$,  at  $z\sim 1$  (assuming that $n$ is not too close to $0$ or not too large). Therefore, the probability distribution $P_n(A|L)$ is expected to have its maximum, of order $D^{1/3}/L^{n+\frac{2}{3}}$, at $A\sim D^{-1/3} L^{n+\frac{2}{3}}$.

Rather than attempting to determine the entire scaling function $F_n(z)$, here we only find its leading-order $z\to 0$ asymptotic. This asymptotic corresponds to the $A\ll D^{-1/3} L^{n+\frac{2}{3}}$ tail of the distribution $P_n(A|L)$ \cite{tooshorttimes}. This large-deviation tail can be obtained by the optimal fluctuation method, akin to geometrical optics. First we need to identify the action functional, corresponding to the Langevin equation (\ref{Langevin}). We start from the probability distribution of a realization of the white Gaussian noise $\xi(t)$ of unit magnitude, see \textit{e.g.} Ref. \cite{Majumdarreview}:
\begin{equation*}
\mathcal{P}[\xi(t)] \sim \exp\left[-\frac{1}{2} \int_0^T \xi^2(t)\, dt\right]\,.
\end{equation*}
Expressing $\xi(t)$ through the particle acceleration $\ddot{x}(t)$ from Eq.~(\ref{Langevin}), we can evaluate the probability distribution of a realization of the random acceleration process in the form of $\sim e^{-S[x(t)]}$ with the action functional
\begin{equation}\label{actiondef}
S[x(t)] = \frac{1}{4 D} \int_0^T \ddot{x}^2(t)\,dt\,.
\end{equation}
The optimal fluctuation method (or geometrical optics) is aimed at finding the ``optimal path" $x_*(t)$ that minimizes functional (\ref{actiondef}) subject to the boundary conditions (\ref{BC1}) and (\ref{BC2}), to the positivity condition $x(t)>0$ for $0<t<T$, and to the integral constraint
\begin{equation}\label{constraint}
I[x(t)] = \int_0^T x^n(t)\,dt = A\,.
\end{equation}
The minimization must be performed not only with respect to different paths $x(t)$, but also with respect to the first-passage time $T$.

Let us rescale the coordinate, $\tilde{x}=x/L$. The action functional (\ref{actiondef}) takes the form
\begin{equation}\label{action10}
S[x(t)]=\frac{L^2}{2D}\,s(\tilde{x}),\quad \text{where} \quad s(\tilde{x}) = \frac{1}{2}\int_0^T \ddot{\tilde{x}}^2(t)\,dt\,.
\end{equation}
The constraint (\ref{constraint}) becomes
\begin{equation}\label{constraint1}
I[\tilde{x}(t)] = \int_0^T \tilde{x}^n (t) dt = \frac{A}{L^n}\,.
\end{equation}
The minimization of the rescaled functional $s(\tilde{x})$ subject to the constraint (\ref{constraint1}) can be achieved by minimizing the modified functional
\begin{equation}\label{actionlambda}
s_{\lambda}[\tilde{x}(t)] = s[\tilde{x}(t)] - \lambda I[\tilde{x}(t)]\,.
\end{equation}
The Lagrange multiplier $\lambda$ turns out to be negative, so we can set $\lambda=-\Lambda^4$, where $\Lambda>0$.
Now we also rescale time, $\tilde{t} = \Lambda t$.  The first-passage time $T$ also gets rescaled, $\tilde{T} =\Lambda T$.
The functional (\ref{actionlambda}) becomes
\begin{equation}\label{actionrescaled}
s_{\lambda}[\tilde{x}(\tilde{t})]= \Lambda^3 \int_0^{\tilde{T}} \left[\frac{\ddot{\tilde{x}}^2(\tilde{t})}{2}+\tilde{x}^n(\tilde{t})\right] d \tilde{t}\,.
\end{equation}
and we will drop the tildes everywhere in the following.
%
Since the rescaled functional $s_0[x(t)]$ (recall that the tildes are dropped) involves the particle acceleration $\ddot{x}(t)$, the Euler-Lagrange equation is of the fourth order (see the Appendix):
\begin{equation}\label{EL}
x^{(4)}(t)+n x^{n-1} (t) = 0,
\end{equation}
where the superscript $(4)$ denotes the fourth derivative with respect to time.  Three boundary conditions for Eq.~(\ref{EL}) come with the formulation of the original stochastic problem, see  Eqs. (\ref{BC1}) and (\ref{BC2}):
\begin{equation}\label{BC3}
x(0)=1\,,\,\, \dot{x}(0)=0\,,\,\, \text{and}\,\,x(T) = 0\,.
\end{equation}
The fourth boundary condition,
\begin{equation}\label{BC4}
\ddot{x}(T)=0\,,
\end{equation}
follows from minimization of the action with respect to all possible variations of the particle velocity $\dot{x}$ at $t=T$ (see the Appendix).

The general solution of the rescaled Euler-Lagrange equation (\ref{EL}) has four arbitrary constants. When this equation is supplemented by the four boundary conditions (\ref{BC3}) and (\ref{BC4}) [and the inequality $x(0<t<T)>0$], the problem of finding the $A\to 0$ asymptotic of  $P_n(A|L)$ is determined completely only for $n=0$ where $A=T$, and one is looking for the distribution $P_n(T|L)$ of first passage times. For all other $n>0$ one should, in addition,  minimize the action $S(A,T)$ with respect to $T$. The minimization yields the \emph{optimal value} of the first-passage time $T=T_*(A)$ which dominates the probability $P_n(A|L)$ that we are after. As we show in the Appendix, this additional minimization brings about a fifth boundary condition
\begin{equation}\label{BC5}
\dddot{x}(T) =0\,.
\end{equation}

Once the optimal path $x(t)$ and, for $n\neq 0$, the optimal value $T=T_*(A)$, are found, we can determine $\Lambda$ from the relation
\begin{equation}\label{Lambda}
\Lambda = \frac{L^n}{A} \int_0^{T_*} x^n(t) \,dt\,,
\end{equation}
which follows from the constraint~(\ref{constraint}) or, equivalently, (\ref{constraint1}). The original action (\ref{actiondef}) can now be written as follows:
\begin{eqnarray}
  S[x(t)] &=&  \frac{L^2 \Lambda^3}{2D} s_0[x(t)]\,,\;\:\text{where}\nonumber\\
  s_0[x(t)]&=&\frac{1}{2} \int_0^{T} \ddot{x}^2(t) dt\,.
  \label{action0}
\end{eqnarray}
Plugging Eq.~(\ref{Lambda}) into the first line of Eq.~(\ref{action0}) we obtain, up to a pre-exponential factor, the $A\to 0$ tail of $P_n(A|L)$. It scales as
\begin{equation}\label{Pscaling}
-\ln P_n(A\to 0|L) \simeq S = \frac{\alpha_n L^{3n+2}}{DA^3}\,,
\end{equation}
where
\begin{equation}\label{alphan}
\alpha_n =\frac{1}{4}\left[\int_0^{T_*}x^n(t) dt\right]^3\int_0^{T_*} \ddot{x}^2(t) dt\,.
\end{equation}
Equation (\ref{Pscaling}) describes a universal essential singularity $\sim \exp (-A^{-3})$ of the $A\to 0$ tail of the distribution. It is much steeper than the essential singularity $\sim \exp (-A^{-1})$ of the first-passage Brownian functionals \cite{MM2020}. 

In fact, the large-deviation scaling (\ref{Pscaling}) (with a yet unknown $\alpha_n$) immediately follows from the exact scaling
(\ref{scaling}) once we realize that the $A\to 0$ asymptotic of the function $F_n(\dots)$ in Eq.~(\ref{scaling}) must exhibit, up to a pre-exponent, the characteristic weak-noise scaling $F_n \sim \exp(-\Phi/D)$, where $\Phi$ depends on $A$ and $L$ but is independent of $D$. Now let us proceed to finding the optimal path, that is to solving Eq.~(\ref{EL}) subject to the boundary conditions (\ref{BC3})-(\ref{BC5}).

\section{Solution}
\label{solution}

\subsection{General}
\label{general}
Equation~(\ref{EL}) is easily solvable for $n=0$, $1$ and $2$, when the equation is linear. We will present these solutions shortly.
In the general case, there is conservation law
\begin{equation}\label{conservation}
\dot{x}(t)\,\dddot{x}(t)-\frac{1}{2}\ddot{x}^2(t)+x^n(t) = C =\text{const},
\end{equation}
which is a higher-order analog of energy conservation in classical mechanics.  The conservation law (\ref{conservation})
reduces the order of Eq.~(\ref{EL}) by one. Using the boundary conditions (\ref{BC3})-(\ref{BC5}) at $t=T$, we find that $C=0$ for all $n>0$ \cite{Emden}.

Evaluating the left hand side of the conservation law~(\ref{conservation}) (where $C=0$) at $t=0$, we uncover one more universal property of the optimal path:
\begin{equation}\label{sqrt2}
\ddot{x}(t=0) =-\sqrt{2}\quad \text{for all} \quad n>0\,.
\end{equation}
Finally, using the conservation law ~(\ref{conservation}) with $C=0$, integration by parts and Eqs.~(\ref{BC3}) and (\ref{BC4}), we can rewrite the expression (\ref{alphan}) for $\alpha_n$ in two equivalent alternative forms:

\begin{equation}\label{alphanalternatives}
\alpha_n =\frac{1}{6}\left[\int_0^{T_*}x^n(t) dt\right]^3 = \frac{27}{32} \left[\int_0^{T_*} \ddot{x}^2(t) dt \right]^4\,.
\end{equation}

\subsection{$n=0$: First-passage time}
\label{fptime}
The first-passage time distribution $P(T|L)$ of the random acceleration process was determined quite some time ago \cite{McKean,Marshall,Burkhardt2014}. Its short-time asymptotic coincides, in the leading order, with the short-time asymptotic of the propagator of the random acceleration. For the zero initial particle velocity,
the exact propagator (see \textit{e.g.} Ref. \cite{Burkhardt2014}) simplifies to
\begin{equation}\label{propagator}
\rho(T,v)=\frac{\sqrt{3}}{2 \pi  D T^2}\, e^{-\frac{3 L^2+3 L T v+T^2 v^2}{ D T^3}}\,,
\end{equation}
where $v=\dot{x}(t=T)$ is the particle velocity (in the original units) at $t=T$. We identify the action, corresponding
to this distribution,
\begin{equation}\label{Srho}
S_{\rho} (T,v) = \frac{3 L^2+3 L T v+T^2 v^2}{D T^3}\,,
\end{equation}
and focus on the large-deviation regime $T\to 0$, where this action is much larger than unity.
Minimizing $S_{\rho} (T,v)$ with respect to $v$, we obtain the optimal value $v_*=-3 L/(2 T)$. The
corresponding minimum of the action,
\begin{equation}\label{Sn=0}
S_{\rho} (T,v_*)=\frac{3 L^2}{4 D T^3}\,,
\end{equation}
determines the small-$A$ asymptotic of $P(T|L)$:
\begin{equation}\label{Pn=0}
-\ln P(T|L) \simeq \frac{3 L^2}{4 D T^3}\,,
\end{equation}
which obeys our asymptotic scaling relation (\ref{Pscaling}) with $\alpha_0=3/4$. Now we will rederive the asymptotic
(\ref{Pn=0}) by using the optimal fluctuation formalism.

For $n=0$ the Euler-Lagrange equation~(\ref{EL}) becomes trivial: $x^{(4)}=0$. Its solution, satisfying the boundary conditions (\ref{BC3}) and (\ref{BC4}),
\begin{equation}\label{x0t}
x(t)= 1-\frac{3 t^2}{2 T^2}+\frac{t^3}{2 T^3}\,,
\end{equation}
is a cubic parabola. Equation~(\ref{Lambda}) yields $\Lambda=1$. Then,  using Eq.~(\ref{action0}), we arrive at Eqs.~(\ref{Sn=0})
and~(\ref{Pn=0}) as to be expected.

\subsection{$n=1$: First-passage area}
\label{fparea}

For $n=1$ the Euler-Lagrange equation~(\ref{EL}) is still very simple: $x^{(4)}=-1$. Its  solution is a quartic parabola. Here we have to demand all five boundary conditions (\ref{BC3})-(\ref{BC5}) which determine the four arbitrary constants and the optimal value of the first-passage time $T_*=2^{3/4}$. The resulting rescaled optimal path,
\begin{equation}\label{x1t}
x(t)= 1-\frac{t^2}{\sqrt{2}}+\frac{t^3}{3 \sqrt[4]{2}}-\frac{t^4}{24}\,,
\end{equation}
is depicted, alongside with the optimal acceleration $\ddot{x}(t)$, in Fig. \ref{n=1,2}.  The optimal acceleration is nothing but the (rescaled) optimal
realization of the white Gaussian noise $\xi(t)$, see Eq.~(\ref{Langevin}). Needless to say, the  optimal realization of the noise looks very differently from a \emph{typical} realization of the noise. Now using Eqs.~(\ref{Pscaling}) and~(\ref{alphan}) for $n=1$, we obtain
\begin{equation}\label{Pn=1}
-\ln P(A|L) \simeq \frac{108 L^5}{625 DA^3}\,,
\end{equation}
with $\alpha_1=108/625$.

\begin{figure}[ht]
  \includegraphics[width=4.2cm]{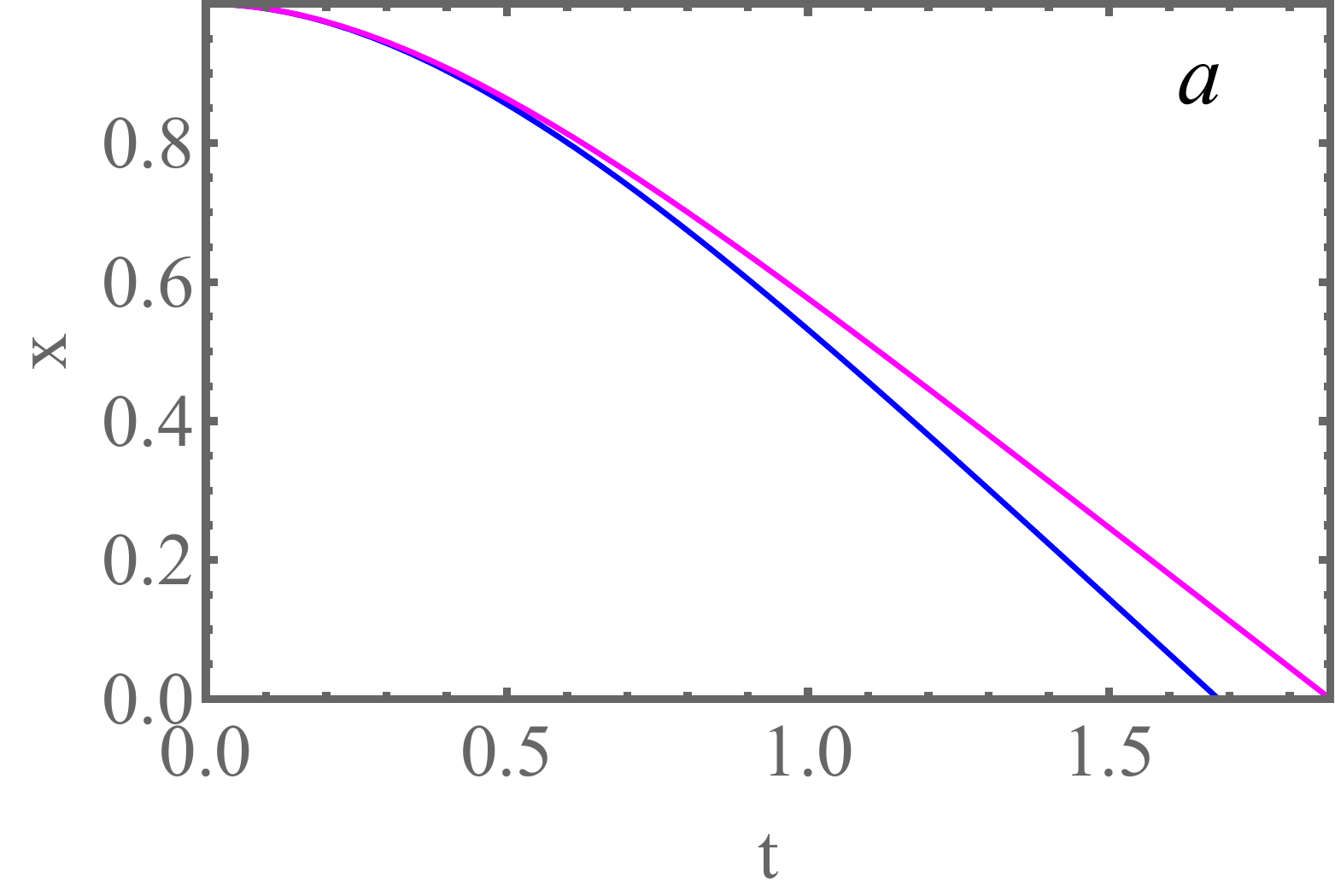}
  \includegraphics[width=4.3cm]{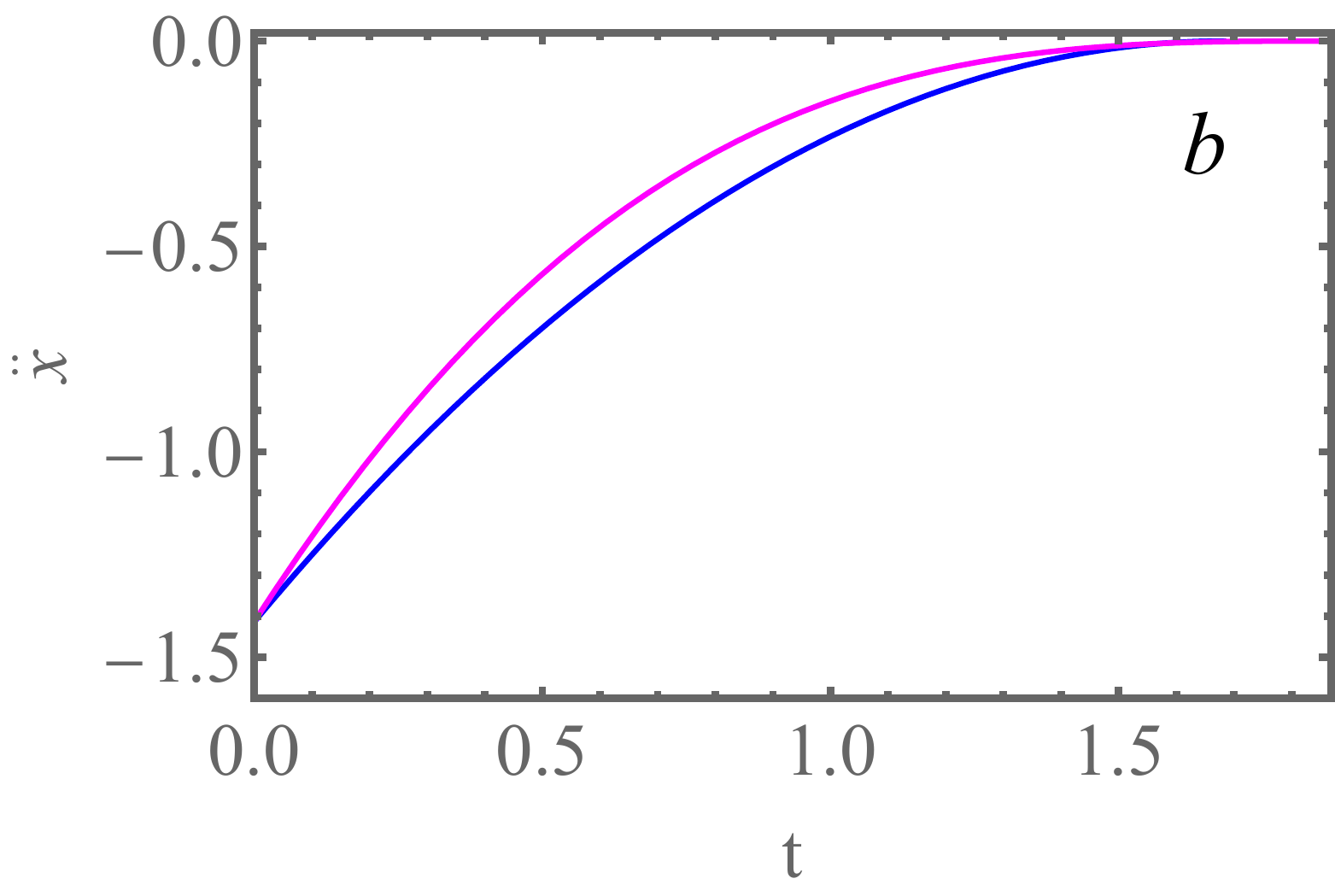}
  \caption{The rescaled optimal path $x(t)$ (a) and optimal acceleration $\ddot{x}(t)$ (b), dominating the $A\to 0$ asymptotics of $P(A|L)$
  for $n=1$ (blue) and $n=2$ (magenta). The optimal first-passage time is $T_*=2^{3/4}$ for $n=1$ and $T_*=\pi/2^{3/4}$ for $n=2$.}
  \label{n=1,2}
\end{figure}

For completeness, we also present $\Lambda$, $T_*$ and  the optimal path $x(t)$ in dimensional variables:
\begin{eqnarray}
\label{x1tdim}
\Lambda&=&\frac{3\cdot 2^{3/4} L}{5 A}\,, \quad T_* = \frac{5A}{3L}\,,\nonumber\\ x(t)&=&L \left(1-2\tau^2+\frac{4\tau^3}{3}-\frac{\tau^4}{3}\right),
\end{eqnarray}
where $\tau=t/T_*$.

\subsection{$n=2$}
\label{squared}
Here the Euler-Lagrange equation (\ref{EL}) is still linear and elementary:
\begin{equation}\label{EL2}
x^{(4)}(t)+2 x(t)=0\,.
\end{equation}
The solution, obeying the boundary conditions (\ref{BC3})-(\ref{BC5}), yields the rescaled optimal path:
\begin{equation}\label{x2t}
x(t)=\frac{\left(1-e^{\frac{t}{T_*}}\right) \sin
   \left(\frac{\pi t}{2 T_*}\right)+\left(1+e^{\frac{t}{T_*}}\right) \cos
   \left(\frac{\pi t}{2 T_*}\right)}{\left(1+e^{-\pi }\right)e^{\frac{\pi t}{2 T_*}} }\,,
\end{equation}
where $T_*=2^{-3/4}\pi$ is the optimal first passage time. Figure \ref{n=1,2} shows this optimal path
alongside with the optimal acceleration $\ddot{x}(t)$.

Using Eqs.~(\ref{Pscaling}) and~(\ref{alphan}) for $n=2$, we obtain
\begin{equation}\label{Pn=2}
-\ln P(A|L) \simeq \frac{27 \tanh^4\left(\frac{\pi }{2}\right)\,L^8 }{256 DA^3}\,,
\end{equation}
Here $\alpha_2=(27/256) \tanh^4\left(\pi/2\right) = 0.074625\dots$.

In the dimensional variables we have
\begin{eqnarray}
\label{x2tdim}
\!\!\Lambda&=&\frac{3 L^2 \tanh\,(\pi/2)}{2^{7/4} A}\,, \quad T_* = \frac{2\pi A \coth (\pi/2)}{3L^2}\,,\nonumber\\
\!\!\frac{x(t)}{L}\!\!&=&\!\!\frac{\left(e^{\pi }\!-\!e^{\pi  \tau }\right) \sin
   \left(\frac{\pi  \tau }{2}\right)\!+\!\left(e^{\pi  \tau }\!+\!e^{\pi }\right) \cos \left(\frac{\pi
    \tau }{2}\right)}{1+e^{\pi }e^{\frac{\pi  \tau }{2}} }\,,
\end{eqnarray}
where $\tau=t/T_*$.

\subsection{Numerics}
\label{numerics}
For arbitrary $n$ the optimal path can be found numerically. We used artificial relaxation in conjunction with iterations over $T$.
Artificial relaxation was implemented as follows. We introduced artificial time $\tau$ and replaced
the Euler-Lagrange equation (\ref{EL}) by the fourth-order partial differential equation
\begin{equation}\label{taueq}
\partial_{\tau} X(t,\tau) = -\partial_{t}^4 X (t,\tau) - n X^{n-1} (t,\tau) \,,
\end{equation}
where the physical time $t$ plays the role of  a coordinate. The sign of the right-hand-side of Eq.~(\ref{taueq}) is chosen
so as to enforce relaxation to a steady-state, $x(t)=X(t,\tau \to \infty)$ which satisfies our Eq.~(\ref{EL}). The initial condition $X(t,\tau=0)$  is chosen qualitatively similar to the expected steady-state solution. Since we do not know the optimal first-passage time $T$ \textit{a priori}, we use iterations.  We first solve Eq.~(\ref{taueq}) with  boundary conditions (\ref{BC3}) and (\ref{BC4}) for a fixed $T$ (the first guess of $T_*$) until the steady-state solution $x(t)$ is reached. Then we evaluate the third derivative $\partial_t^{(3)}X(t,\tau\gg 1)$ at $t=T$, and iterate $T$ until the third derivative vanishes [as Eq.~(\ref{BC5}) demands] with desired accuracy. Alternatively, one can iterate until $\partial_t^{2}X(t,\tau\gg 1)$ at $t=0$ approaches $-\sqrt{2}$, see Eq.~(\ref{sqrt2}). We validated the method by comparing the numerically found $x(t)$ with the analytical solutions for $n=1$ and $2$.
The accuracy was monitored by checking the conservation law~(\ref{conservation}) with $C=0$. Once $T_*$ and $x(t)$ are found, we can evaluate $\alpha_n$ from any of the equations~(\ref{alphan}) or (\ref{alphanalternatives}). We used a standard PDE solver of ``Mathematica" \cite{wolfram}.

\begin{figure}[ht]
  \includegraphics[width=4.1cm]{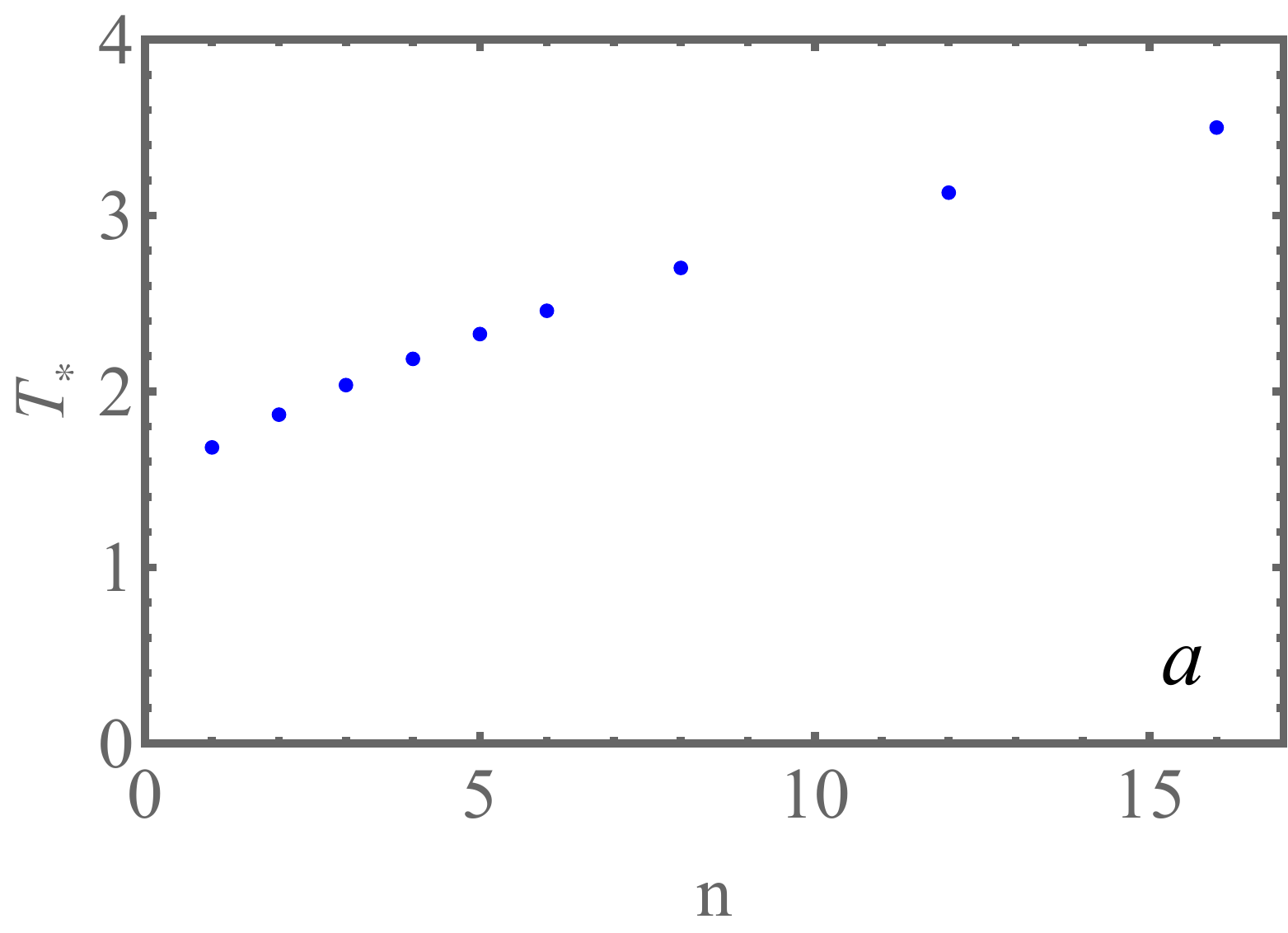}
  \includegraphics[width=4.4cm]{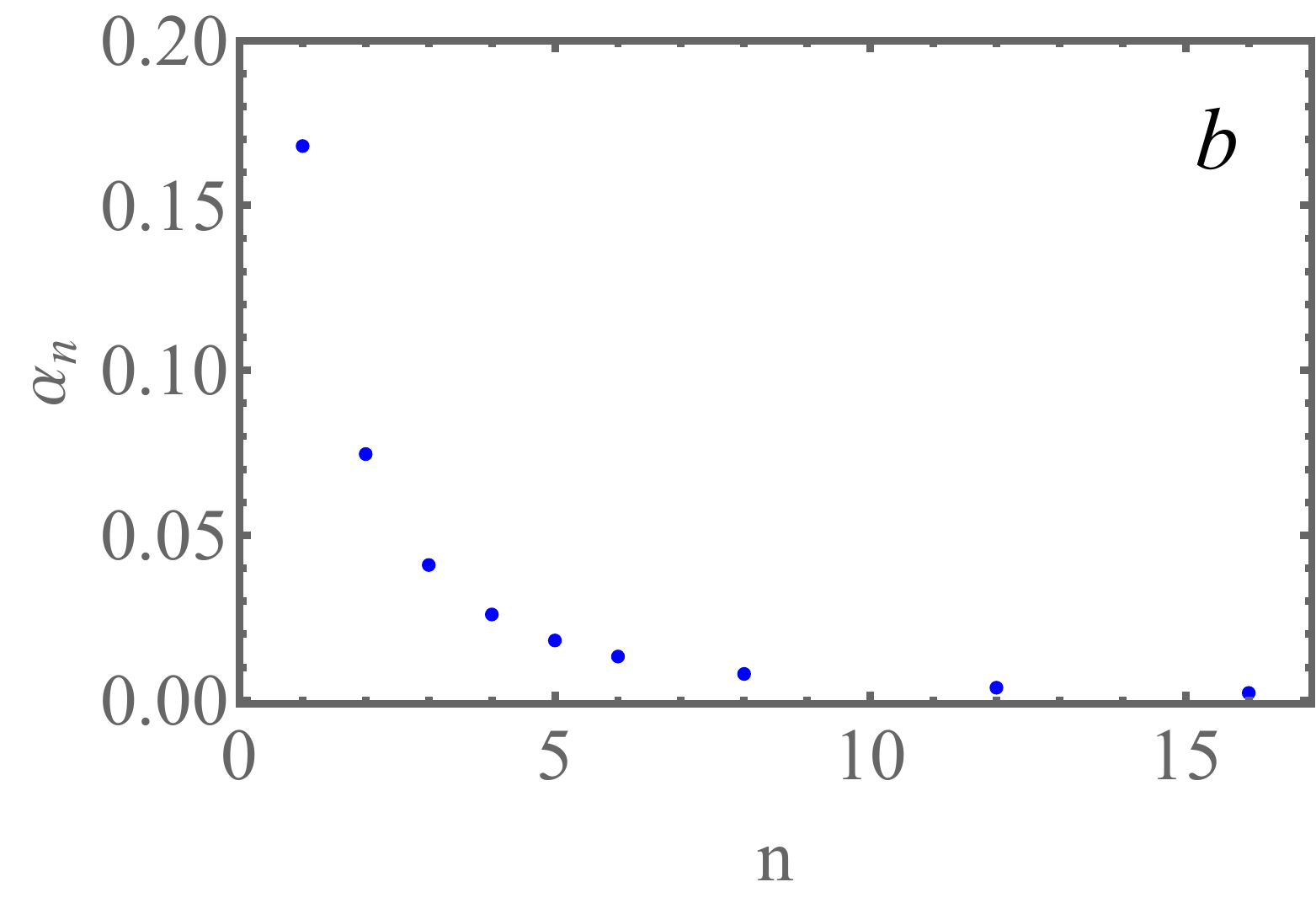}
  \caption{The optimal first-passage time $T_*$, conditioned on Eq.~(\ref{constraint0}) (a), and the factor $\alpha_n$ which enters Eqs.~(\ref{Pscaling}) and (\ref{alphan}) (b) are plotted as functions of $n$. The points $n=1$ and $n=2$ were obtained analytically, the rest of the points numerically.}
  \label{listplots}
\end{figure}

Figure~\ref{n=3,4} shows the numerically found optimal paths $x(t)$
and the optimal accelerations $\ddot{x}(t)$ for $n=3$ and $4$. In these cases $T_*\simeq 2.036$ and $2.185$, respectively, whereas the $A\to 0$ asymptotics of $P(A|L)$ are described by Eq.~(\ref{Pscaling}) with $\alpha_3\simeq 0.041$
and $\alpha_4 \simeq 0.026$. Overall, we solved the problem numerically and found  the optimal first passage time $T_*$ and the factor $\alpha_n$ for a range of $n$, see Fig. \ref{listplots}. As one can see, $T_*$ increases with $n$, while $\alpha_n$ decreases.

\begin{figure}[ht]
  \includegraphics[width=4.2cm]{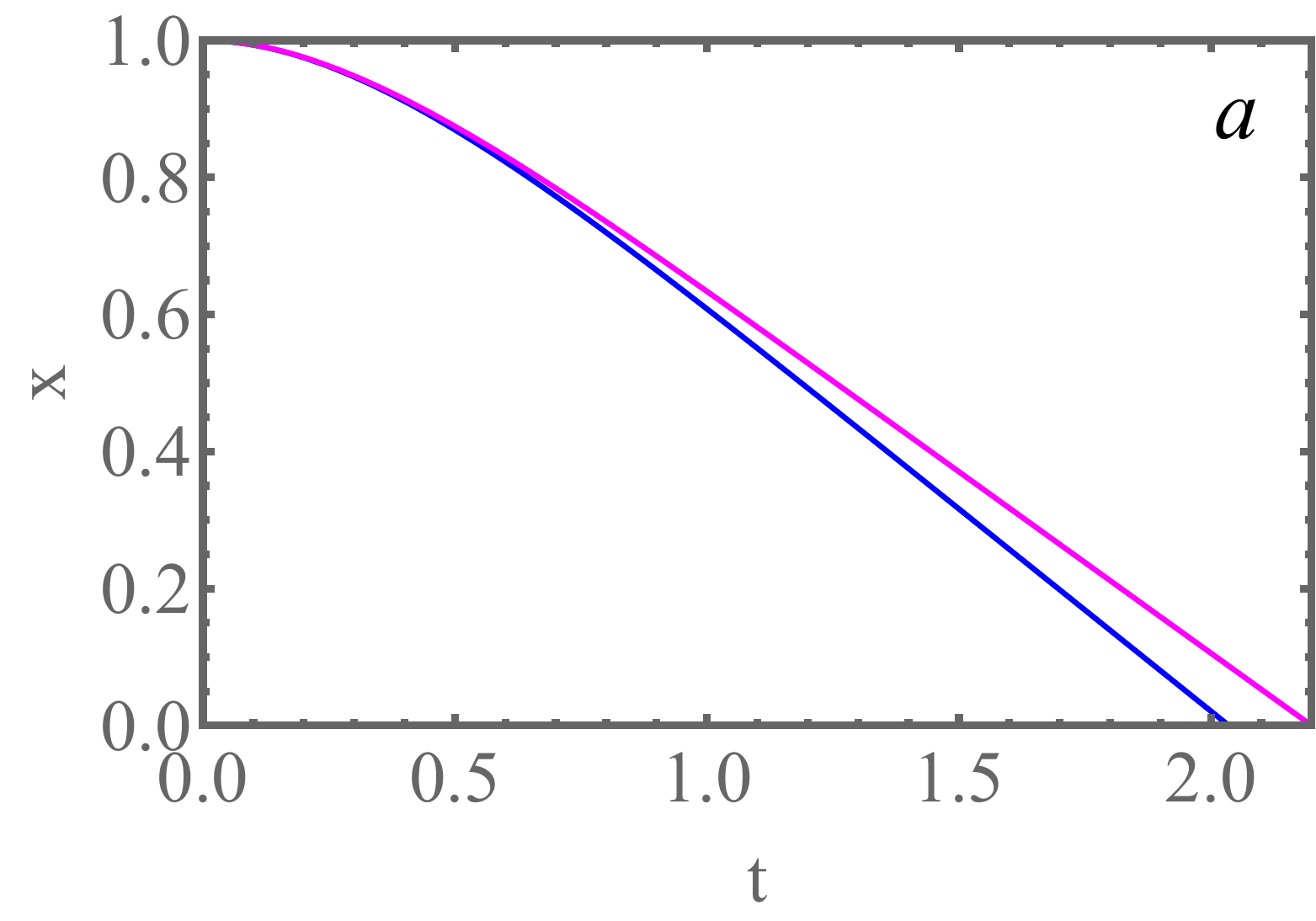}
  \includegraphics[width=4.3cm]{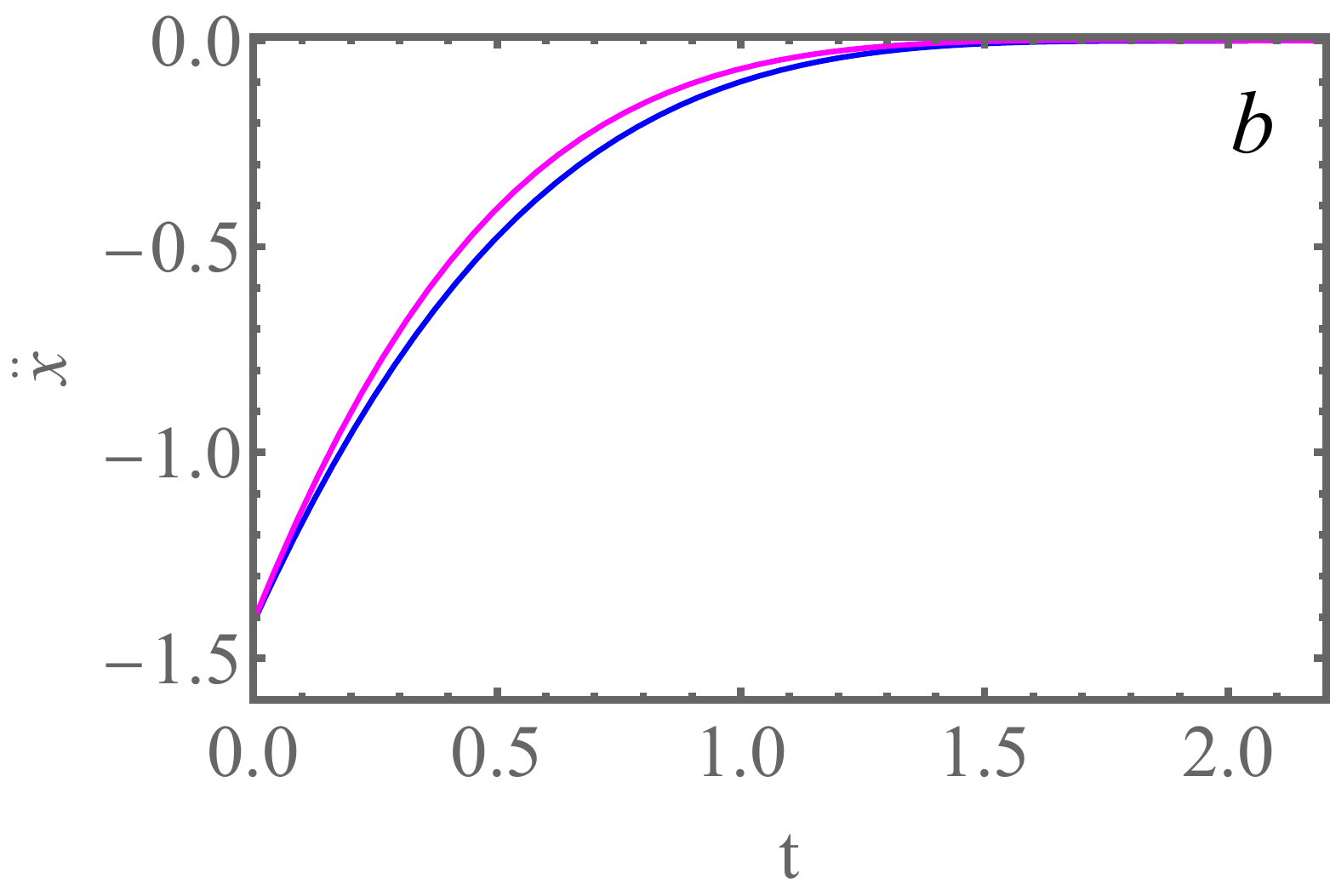}
  \caption{Numerically found rescaled optimal path $x(t)$ (a) and optimal acceleration $\ddot{x}(t)$ (b), dominating the $A\to 0$ asymptotics of $P(A|L)$
  for $n=3$ (blue) and $n=4$ (magenta). The optimal first-passage time is $\simeq 2.036$ for $n=3$ and $T_*\simeq 2.20$ for $n=4$.}
  \label{n=3,4}
\end{figure}

\section{Summary and Discussion}
\label{discussion}

Statistics of first-passage functionals provide a useful characterization of random processes. Here we evaluated the $A\to 0$ tail of these statistics for the random acceleration process. We also used this problem to extend the optimal fluctuation (or geometrical optics) method to a stochastic process of a higher-order. In addition to the $A\to 0$ asymptotic of the probability distribution $P_n(A|L)$, we calculated analytically and numerically the optimal paths of the conditioned processes at different $n$. These provide an interesting insight into the nature of large deviations in this system.

The geometrical-optics calculations can be extended to the case where the particle velocity at $t=0$ is nonzero: $v_0\neq 0$. The more interesting case here is $v_0<0$, when the particle arrives at the origin with probability 1. Here a new effect appears: the particle can reach the origin \emph{deterministically} along the ballistic trajectory $x(t) = L-|v_0| t$. The expected  first-passage time (not subject to any additional constraint) is $\bar{T}=L/|v_0|$.  Evaluating the functional $I[x(t)]$ in Eq.~(\ref{constraint}) on the ballistic trajectory, we obtain
\begin{equation}\label{Abar}
A\equiv \bar{A} = \frac{L^{n+1}}{(n+1)|v_0|}\,.
\end{equation}
If we condition the process on $A<\bar{A}$ or $A>\bar{A}$, the optimal path will be  different from the deterministic one, and it can be found with the same formalism we used, except that the second condition in Eq. (\ref{BC3}) should be replaced by $\dot{x}(t=0)=-|v_0|$
(in the dimensional variables). The resulting action~(\ref{actiondef}) vanishes at $A=\bar{A}$. In a small vicinity around $A=\bar{A}$ the action is quadratic in $A-\bar{A}$. For such small fluctuations of $A$ the probability distribution is approximately Gaussian. Applicability of geometrical optics requires that the resulting action be much larger than unity.

The simplest example of such a calculation is the evaluation
of the first-passage time distribution itself, that is $n=0$.  Here instead of Eq.~(\ref{x0t}) we obtain
\begin{equation}\label{xtn=0}
x(t)=L+v_0 t-\frac{3(L+v_0T) t^2}{2 T^2}+\frac{(L+v_0T)t^3}{2 T^3}\,.
\end{equation}
The resulting probability density $P(T,L,v_0<0)$ is given,  up to a pre-exponential factor, by the expression
\begin{equation}\label{actionv0}
-\ln P(T,L,v_0<0) \simeq S= \frac{3 (L-|v_0|T)^2}{4 D T^3}\,,
\end{equation}
which agrees  with the short-time asymptotic of the exact propagator of the random acceleration for nonzero initial particle velocity \cite{Burkhardt2014}.
As to be expected, the action~(\ref{actionv0}) vanishes at $T=\bar{T}=L/|v_0|$. For $v_0=0$ Eq.~(\ref{actionv0}) coincides with Eq.~(\ref{Sn=0}), again as to be expected.

Finally, the problem of statistics of the first-passage functionals $I[x(t)] = \int_0^T x^n(t) dt$ can be extended to a whole family
of processes, described by the Langevin equation $d^kx(t)/dt^k = \sqrt{2D} \xi(t)$, where $k$ is any positive integer. The cases of $k=1$ and $k=2$ correspond to the Brownian motion and random acceleration, respectively. Again, let
$x(0)=L$, and again suppose for simplicity that all the derivatives of $x(t)$ with order less than $k$ vanish at $t=0$. Then the exact scaling behavior of probability distribution $P_n^{(k)}(A|L)$ of the values $I[x(t)]=A$ follows from dimensional analysis:
\begin{equation}\label{scalingk}
P^{(k)}_n(A|L) = \frac{D^{\nu}}{L^{n+2\nu}}\,F^{(k)}_n\left(\frac{D^{\nu} A}{L^{n+2\nu}}\right)\,,
\end{equation}
where $F^{(k)}_n(z)$ is an unknown scaling function, and $\nu=1/(2k-1)$.
In its turn, the leading-order $A\to 0$ asymptotic of $P_k(A|L)$ must have the characteristic weak-noise form
\begin{equation}\label{Pscalingk}
-\ln P_n^{(k)}(A\to 0|L) \simeq \frac{\alpha_n^{(k)} L^{\frac{n}{\nu}+2}}{DA^{1/\nu}}\,,
\end{equation}
where $\alpha_n^{(k)}$ is a numerical factor which depends on $k$ and $n$. As one can see from Eq.~(\ref{Pscalingk}), for all these models theory predicts am essential singularity at $A\to 0$, and the singularity becomes stronger as $k$ is increased.

\section*{Acknowledgments}

The author is very grateful to Satya N. Majumdar for a useful discussion. This research was supported by the Israel Science Foundation (Grant No. 1499/20).

\vspace{0.5 cm}

\appendix


\section{Derivation of Eq.~(\ref{EL}) and boundary condition (\ref{BC5}).}
\label{appendix}
\noindent
Here we temporarily switch back to the original variables and consider a linear variation of the constrained action functional
\begin{equation}\label{actionapp}
s_{\lambda}[x(t),T] =\int_0^T \left(\frac{\ddot{x}^2}{2}-\lambda x^n\right) dt
\end{equation}
with respect to small variations of both $x(t)$ and $T$:  $x(t) \to x(t)+\delta x(t)$ and
$T\to T +\delta T$.  We need to linearize the variation
\begin{equation}\label{A1}
\delta s_{\lambda} = s[x(t)+\delta x(t), T+\delta T] -s[x(t), T]
\end{equation}
with respect to $\delta x$ and $\delta T$.  The linearization yields, after simple algebra,
\begin{eqnarray}
  \delta s_{\lambda}  &=& \int_0^T \left( \ddot{x} \delta \ddot{x}-\lambda n x^{n-1} \delta x \right) dt \nonumber \\
  &+& \int_T^{T+\delta T} \left(\frac{\ddot{x}^2}{2}-\lambda x^n\right)dt\,. \label{A2}
\end{eqnarray}
Performing two integrations in parts in the first integral, evaluating the second integral in the limit of $\delta T\to 0$, and taking into account the boundary conditions (\ref{BC3}), we obtain
\begin{eqnarray}
  \delta s_{\lambda}  &=&\int_0^T\left(x^{(4)}-\lambda n x^{n-1}\right) \delta x\,dt+\ddot{x}(T) \delta \dot{x}(T) \nonumber \\
  &+& \left[\frac{\ddot{x}^2(T)}{2}-\lambda x^n(T)-\dddot{x}(T)\dot{x}(T)\right]\delta T\,. \label{A3}
\end{eqnarray}
Each of the three terms in the variation must vanish independently for arbitrary $\delta x$ and $\delta T$. The first term yields
the Euler-Lagrange equation $x^{(4)}-\lambda n x^{n-1}=0$
which, upon the rescaling $\Lambda t \to t$ (we recall that $\lambda \equiv -\Lambda^4$), coincides with Eq.~(\ref{EL}).
The second term yields the boundary condition (\ref{BC4}). Using the latter condition and the condition $x(T)=0$ in the third term, we arrive at the condition $\dddot{x}(T)\dot{x}(T)=0$. Now we have to choose one of two options: $\dot{x}(T)=0$ or $\dddot{x}(T)=0$ (they cannot hold simultaneously, as the problem then would be overdetermined). As one can check \cite{minmax}, the condition $\dot{x}(T)=0$ would give a local \emph{maximum} of the action functional $S[x(t)]$ as a function of $T$, whereas the condition $\dddot{x}(T)=0$ yields the desired minimum.



\begin{thebibliography}{99}
\bibitem{tooshorttimes} At very short times, the random acceleration model (where a real stochastic force is modelled by white noise) breaks down, and some additional, ``microscopic" physics should be taken into account. Therefore, as it often occurs in physics, our small-$A$ results should be considered as an intermediate asymptotic: they are expected to be applicable when $A$ is much smaller than $D^{-1/3}L^{n+2/3}$,  but much larger than a (system-specific) microscopic time scale where the random-acceleration model breaks down.
\bibitem{Emden} By introducing a new function $z(x) = [-\dot{x}(t)]^{3/2}$, Eq.~(\ref{conservation}) can be reduced to a second-order equation
$$
z^{\prime\prime}(x)= -\frac{3}{2}x^n z^{-5/3}(x)\,,
$$
which is known as the Emden-Fowler equation \cite{Polyanin}. Unfortunately, its exactly soluble cases do not seem to extend beyond $n=0,1$ and $2$, where it is much easier to solve Eq. (\ref{EL}) directly.
\bibitem{minmax} We did it analytically for $n=1$ and $2$, and numerically for several other $n$.
\end{thebibliography}
\end{document}